\documentclass[twocolumn,tighten]{aastex631}
\usepackage{graphicx}
\usepackage{amsmath}
\usepackage{natbib}
\usepackage{amssymb}
\usepackage{ulem}
\usepackage{sidecap}
\usepackage{hyperref}

\def\MP{\ensuremath{m_{\rm{p}}}}

\def\BC{\ensuremath{k_{\rm{B}}}}

\def\Msun{\ensuremath{\mathrm{M}_\odot}}

\def\Rb{\ensuremath{R_{\rm B}}}

\shorttitle{Late-time Radio in TDEs}
\shortauthors{Matsumoto \& Piran}

\begin{document}

\title{Late-time Radio Flares in Tidal Disruption Events}

\author[0000-0002-9350-6793]{Tatsuya Matsumoto}
\affil{Department of Astronomy, Kyoto University, Kitashirakawa-Oiwake-cho, Sakyo-ku, Kyoto, 606-8502, Japan}
\affil{Hakubi Center, Kyoto University, Yoshida-honmachi, Sakyo-ku, Kyoto, 606-8501, Japan}
\author[0000-0002-7964-5420]{Tsvi Piran}
\affil{Racah Institute of Physics, The Hebrew University of Jerusalem, Jerusalem, 91904, Israel}

\begin{abstract}
Radio monitoring unveiled late (hundreds to a thousand days) radio flares in a significant fraction of tidal disruption events. We propose that these late-time radio flares are a natural outcome if the surrounding density profile flattens outside the Bondi radius. At the Bondi radius, the outflow is optically thin (above a few GHz) to synchrotron self-absorption. As more and more material is swept up, the radio emission rises asymptotically as $\propto t^3$ until the outflow begins to decelerate. A Detection of such a rise and a late-time maximum constrains the black hole mass and the mass and energy of the radio-emitting outflow. We show that this model can give reasonable fits to some observed light curves, leading to reasonable estimates of the black hole and outflow masses. We also find that the slope of the density profile within the Bondi radius determines whether an early-time ($\sim10^2\,\rm days$) radio peak exists.
\end{abstract}


\section{Introduction}
\label{sec:introduction}

Tidal disruption events \citep[TDEs,][]{Hills1975,Rees1988} by supermassive black holes (SMBHs) produce, in addition to optical/UV and X-rays, radio emission. The radio emission is powered by synchrotron from an outflow interacting with circum-nuclear medium (CNM) \citep[see][for a review]{Alexander+2020}. Usually, TDEs' radio emission have been discovered by prompt radio follow-ups of optical/X-ray TDEs at $\sim100\,\rm days$ after the discovery, with a typical luminosity below $\lesssim10^{39}\,\rm erg\,s^{-1}$. These radio signals are produced by non-relativistic outflows \citep{Krolik+2016,Alexander+2016,Yalinewich+2019b,Anderson+2020,Stein+2021,Cendes+2021b}. A small fraction of TDEs that launch a relativistic jet produce a very bright radio emission \citep{Zauderer+2011,Berger+2012,Cenko+2012,Zauderer+2013,Pasham+2015,BrownGC+2017,Eftekhari+2018,Cendes+2021,Andreoni+2022,Pasham+2022,Rhodes+2023}.

Following the surprising discovery of late radio emission from ASASSN-15oi \citep{Horesh+2021}, numerous radio flares as late as $\sim1000\,\rm days$ after the optical discoveries \citep{Horesh+2021b, Cendes+2022b,Goodwin+2022,Perlman+2022,Sfaradi+2022,Cendes+2023,Goodwin+2023b,Somalwar+2023c,Sfaradi+2024,ZhangFabao+2024,Christy+2024} have been observed. Recently \cite{Cendes+2023} reported that about $40\,\%$ of optical TDEs are accompanied by late-time radio flares. Such late-time emission is difficult to produce by a conventional scenario in which  a radio-emitting outflow is ejected at around the same time as the stellar disruption, and several  scenarios  such as late-time launch of outflows due to delayed formation of an accretion disk \citep[for disk winds]{Cendes+2023} or misalignment of disk and BH spin \citep[for relativistic jets]{Teboul&Metzger2023,Lu+2023} 
have been proposed. 

\cite{Matsumoto&Piran2023} proposed that the fast-rising late-time radio flare in AT 2018hyz \citep{Cendes+2022b,Sfaradi+2024} can be explained by an off-axis relativistic jet in which initially the radio signal is beamed away from the observers. The flux rapidly increases when the jet decelerates and its beam moves into the line of sight. However, some of the events reported in \cite{Cendes+2023}, e.g.,  AT2019dsg, do not satisfy the criterion for which the off-axis scenario holds \citep[see][]{Matsumoto&Piran2023}. Moreover, given the event rate of on-axis jetted TDEs $\simeq0.01-0.1\,\rm Gpc^{-3}\,yr^{-1}$ \citep[e.g.,][]{Andreoni+2022}, which is about $\sim10^4$ times less than optical/X-ray TDEs $\simeq10^3\,\rm Gpc^{-3}\,yr^{-1}$ \citep{VanVelzen2018,Sazonov+2021,Yao+2023} it is unlikely, that all the observed late flares arise from off-axis jets.

Here, we propose an alternative model that does not employ a relativistic jet or a delayed outflow. In our scenario, the late-time radio flares arise due to a transition in the density profile of the surrounding medium from gradually declining to a constant at larger distances. The observed late-time radio is an optically thin synchrotron emission from an outflow traveling into this constant-density medium. While near the SMBH the density profile decreases, a constant density profile is naturally expected outside of the Bondi radius \citep{Bondi1952}. 

The outline of the paper is as follows. In Sec.~\ref{sec:method}, we describe our model for radio light curves, showing that it can explain the diversity of the radio light curve (Sec.~\ref{sec:diversity}) and demonstrating that {it can reproduce } some of the observed events (Sec.~\ref{sec:fit}). We also discuss estimating the BH mass and outflow's mass within our model (Sec.~\ref{sec:parameter}). We summarize our results in Sec.~\ref{sec:summary}.

\section{The Radio Light Curve}\label{sec:method}
The radio-producing outflow initially expands into a CNM, described by a single power-law function. The main difference from previous studies focusing on early-time flares \citep[e.g.,][]{Krolik+2016,Alexander+2016,Anderson+2020} is that we consider a decreasing density profile that is smoothly connected to interstellar medium (ISM) with a constant density at the Bondi radius:
\begin{align}
R_{\rm B}&=\frac{GM_{\bullet}}{c_{\rm s}^2}\simeq5.8\times10^{16}{\,\rm cm\,}M_{\bullet,6}T_7^{-1}\ ,
	\label{eq:R_bondi}
\end{align}
where $G$ is the gravitational constant and $M_{\bullet}$ is the mass of the SMBH. The sound velocity is given by $c_{\rm s}=\left(\gamma\BC T/\mu \MP\right)^{1/2}\simeq480{\,\rm km\,s^{-1}\,}T_7^{1/2}$, where $\gamma=5/3$, $\BC$, $T$, $\mu=0.6$, and $\MP$ are the adiabatic index, Boltzmann constant, ISM temperature, mean molecular weight, and proton mass, respectively. For the BH mass, we use the normalization of $M_{\bullet}=10^6M_{\bullet,6}\Msun$. For the ISM temperature, we adopt a notation $Q=Q_x10^x$ in the cgs unit, which is also used hereafter for other quantities unless specified. The fiducial value of the temperature that we use, $T=10^7\,\rm K$, is much higher than that of typical ISM \citep[warm neutral medium, e.g.,][]{Draine2011}. However, this is indeed an observationally motivated value. For Sgr A*, \textit{Chandra} detected a diffuse keV  X-ray emission across a sub-parsec scale in our galactic center, corresponding to the Bondi radius scale \citep{Baganoff+2003}. This emission comes from shock-dissipated stellar winds from massive stars orbiting around the SMBH \citep[e.g.,][]{Paumard+2006,Genzel+2010}.\footnote{Theoretical works modeling the accretion of stellar wind onto Sgr A* also suggest that the density profile flattens at the radius where an inflow and outflow separates \citep[e.g.,][]{Quataert2004,Generozov+2017,Ressler+2018}, the so-called the stagnation radius.} M87 and other nearby galaxies' core show a similar keV X-ray emission across the Bondi radius \citep{DiMatteo+2003,Ho2008}. 

The actual density profile around the galactic center is not well determined. A constant density outside of the Bondi radius is a minimal assumption, which is justified because the density profile is expected to connect to the galactic ISM at some distance from the SMBH. With this assumption we avoid introducing additional undetermined parameters. The profile could be more complicated. For example, if the source of the
accreting materials onto the SMBH is winds from orbiting stars, the density profile depends on the stellar distribution. For Sgr A* the profile would be slowly decreasing with the radius \citep[e.g.,][]{Quataert2004, Ressler+2018}. Alternatively an increasing profile within a some radial range has been suggested \citep{Cuadra+2015}. It should be noted that, as shown below, the second peak in the radio light curve should appear as long as the profile has a slope less than $\lesssim1.6$ and it is not strongly dependent on this assumption (of course, the detailed shape of the light curve will depend on the exact density profile).

\begin{figure}
\begin{center}
\includegraphics[width=85mm, angle=0, bb=0 0 276 215]{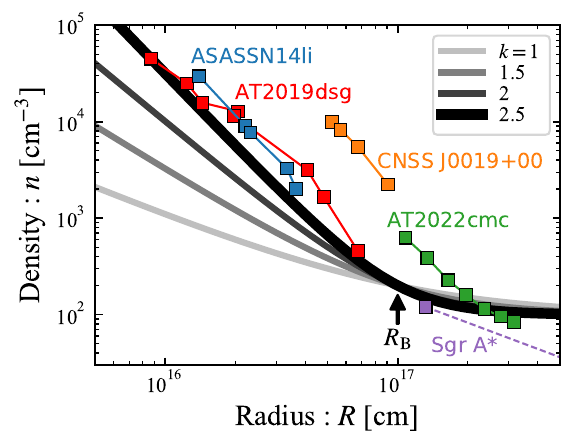}
\caption{Density profiles for different slopes of the CNM components $k$ and a fixed Bondi radius at $\Rb=10^{17}\,\rm cm$ (see Eq.~\ref{eq:profile}). The reconstructed profiles shown are obtained for the observed data points are for ASASSN14li \citep{Krolik+2016,Alexander+2016}, CNSS J0019+00 \citep{Anderson+2020}, AT2019dsg \citep{Stein+2021,Cendes+2021b,Matsumoto+2022}, AT2022cmc \citep{Matsumoto&Metzger2023}, and Sgr A* \citep{Baganoff+2003}.}
\label{fig:profile}
\end{center}
\end{figure}

The combined density profile is described by:
\begin{align}
n(R)&=n_{\rm ISM}\left[\left(\frac{R}{\Rb}\right)^{-k}+1\right]\ ,
	\label{eq:profile}
\end{align}
where the first term corresponds to a decreasing power-law profile with a slope $k$, representing the CNM. Analyses of radio TDEs roughly constrain the slope $k\simeq2.5$ \citep[e.g.,][]{Alexander+2020}. Fig.~\ref{fig:profile} depicts the density profile for various slopes and a fixed Bondi radius at $10^{17}\,\rm cm$ along with profiles reconstructed for observed events.

\begin{table}
\begin{center}
\caption{Model parameters and their fiducial values for the synchrotron light curve.}
\label{table:parameter}
\begin{tabular}{lll}

\hline
$p$&Slope of electron energy distribution&$2.5$\\
$\varepsilon_{\rm e}$&Energy fraction of non-thermal electron&$0.1$\\
$\varepsilon_{\rm B}$&Energy fraction of magnetic field&$0.01$\\
$M_{\rm ej}$&Outflow mass&$0.1\,\Msun$\\
$\beta_0$&Initial velocity&$0.1$\\
$n_{\rm ISM}$&ISM density&$100\,\rm cm^{-3}$\\
$\Rb$&Bondi radius&$10^{17}\,\rm cm$\\
$k$&Slope of the CNM density profile&2.5\\
\hline
\end{tabular}
\end{center}
\end{table}

Radio light curves are calculated by a standard formalism, i.e., a non-relativistic version of \cite{Sari+1998}. We adopt the same numerical code used in our previous studies \citep{Ricci+2021,Bruni+2021,Matsumoto&Piran2021b,Matsumoto&Metzger2023}. This model has several parameters: The power-law index of the electron distribution $p$ and the relativistic electrons and magnetic field energy fractions at the post-shock region $\varepsilon_{\rm e}$ and $\varepsilon_{\rm B}$. The ejecta is assumed to be a quasi-spherical non-relativistic outflow, characterized by an initial velocity $\beta_{\rm in}$, mass $M_{\rm ej}$, and an opening angle $\Omega$. These parameters, as well as the parameters characterizing the density profile, are listed in Table~\ref{table:parameter}. The unbound debris,  an inevitable outcome of any TDE, is the most natural source of the outflow \citep{Krolik+2016,Yalinewich+2019b,Matsumoto&Piran2021b}. Additional sources such as disk wind \citep{Strubbe&Quataert2009,Metzger&Stone2016,Dai+2018} or collision between bound debris \citep{Lu&Bonnerot2020} have been proposed. Here, we do not specify a specific scenario for the origin of the outflow. We characterize the outflow using its total mass, $M_{\rm ej}$, velocity, $\beta_{\rm in}$, and solid angle, $\Omega$. For simplicity, we assume that all the matter moves at the same velocity. The model can be easily generalized to include the effect of a velocity distribution, which is important during the deceleration phase.

\begin{figure}
\begin{center}
\includegraphics[width=85mm, angle=0,bb=0 0 285 211]{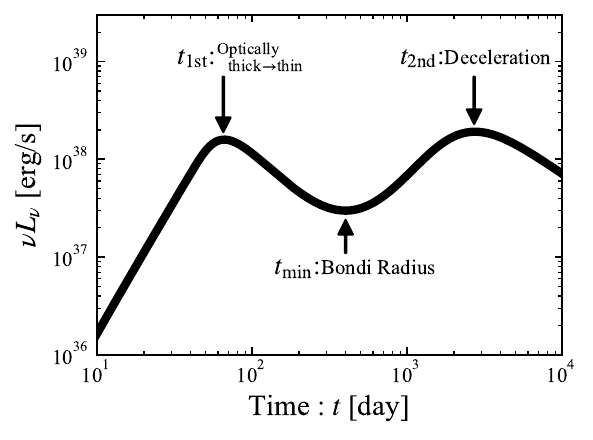}
\caption{A radio light curve at a single observed frequency, $6\,\rm GHz$, calculated for the fiducial parameters (see  Table~\ref{table:parameter}). Initially, the outflow is optically thick to SSA, and the light curve rises until SSA becomes negligible at $t_{\rm 1st}$. After the first peak, the emission is optically thin, and the light curve declines, reaching a minimum when the outflow reaches the Bondi radius at $t_{\rm min}$. The second peak is caused when the deceleration happens at $t_{\rm 2nd}$. This double-peak structure is consistent with the observations of AT2019dsg and AT2020vwl (see Fig.~\ref{fig:lc_obs}).
}
\label{fig:lc}
\end{center}
\end{figure}

Fig.~\ref{fig:lc} depicts the resulting light curve at $\nu=6\,\rm GHz$ for our fiducial parameters listed in Table~\ref{table:parameter}. Remarkably, the light curve has two peaks that are qualitatively similar to those found in AT2019dsg and AT2020vwl (see Fig.~\ref{fig:lc_obs}). The first peak is caused by a transition from optically thick for synchrotron self-absorption (SSA) to an optically thin state, which is common for radio emission in TDEs. For a steep CNM density profile, the optically thin radio light curve declines until the freely expanding outflow reaches the Bondi radius. Beyond this radius, the density profile flattens and the light curve rises because the outflow sweeps up more and more electrons. When the outflow sweeps up a comparable amount of mass to its original mass, it decelerates, and the light curve begins declining again.

The behavior of the radio light curve can be discussed more quantitatively. The synchrotron luminosity for optically thick and thin cases are estimated by:\footnote{The formulae for the synchrotron flux, Eqs.~\eqref{eq:L_thick} and \eqref{eq:L_thin}, and SSA frequency Eq.~\eqref{eq:nu_a} are given in \cite{Matsumoto&Piran2021b} (corresponding equations are Eqs.~12, 11, and 8). At the same time, we update the estimate of the total number of electrons from $\Omega n R^3$ to $\Omega n R^3/3$, which is useful for profiles flattening at a larger radius.}

\begin{align}
\left(\nu L_{\nu}\right)_{\rm thick}&\overset{p=2.5}{\simeq}1.8\times10^{40}{\,\rm erg\,s^{-1}\,}
	\nonumber\\
&\,\,\,\,\,\,\varepsilon_{\rm B,-2}^{-1/4}n_2^{-1/4}R_{17}^2\beta_{-1}^{-1/2}\nu_{\rm 6GHz}^{7/2}\left(\frac{\Omega}{4\pi}\right)\ ,
    \label{eq:L_thick}\\
\left(\nu L_{\nu}\right)_{\rm thin}&\overset{p=2.5}{\simeq}1.8\times10^{36}{\,\rm erg\,s^{-1}\,}
	\nonumber\\
&\,\,\,\,\,\,\bar{\varepsilon}_{\rm e,-1}\varepsilon_{\rm B,-2}^{\frac{p+1}{4}}n_2^{\frac{p+5}{4}}R_{17}^3\beta_{-1}^{\frac{p+5}{2}}\nu_{\rm 6GHz}^{\frac{3-p}{2}}\left(\frac{\Omega}{4\pi}\right)\ ,
    \label{eq:L_thin}
\end{align}
where $\Omega$ is the solid angle of the outflow, $\beta$ is the outflow velocity normalized by the speed of light, and we define $\bar{\varepsilon}_{\rm e}\equiv 4\left(\frac{p-2}{p-1}\right)\varepsilon_{\rm e}$. We normalize the observer frequency by 6 GHz. Hereafter, the numerical values, indicated by the notation $\overset{p=2.5}{\simeq}$, are evaluated for $p=2.5$. It should be noted that in these formulae, the total number of electrons is approximated by $N(R)=\Omega n(R)R^3/3$, where $n(R)$ is the number density at the shock front $R$. However, for the profile of Eq.~\eqref{eq:profile}, the number is given by
\begin{align}
N(R)=\frac{\Omega}{3}n_{\rm ISM}R^3\left[\frac{3}{3-k}\left(\frac{R}{\Rb}\right)^{-k}+1\right]\ .
    \label{eq:Ntot}
\end{align}
Therefore, our approximation results in an underestimate of the luminosity for Eqs.~\eqref{eq:L_thick} and \eqref{eq:L_thin} at $R\lesssim\Rb$ by a factor of ${3-k}/{3}$.

For a CNM density profile of $n\propto R^{-k}$ and freely-expanding outflow ($R\propto t$), the luminosity evolves with time as  
\begin{align}
\left(\nu L_{\nu}\right)_{\rm thick}&\propto t^{\frac{k+8}{4}}\ ,
    \label{eq:L_thick2}\\
\left(\nu L_{\nu}\right)_{\rm thin}&\propto t^{\frac{12-k(p+5)}{4}}\ .
    \label{eq:L_thin2}
\end{align}
For density slopes with $k>\frac{12}{p+5}\overset{p=2.5}{=}1.6$, the light curve at a given frequency $\nu$, has a peak when the SSA frequency, $\nu_{\rm a}$, reaches $\nu$. The SSA frequency is given by 
\begin{align}
\nu_{\rm a}\overset{p=2.5}{\simeq}0.35{\,\rm GHz\,}\bar{\varepsilon}_{\rm e,-1}^{\frac{2}{p+4}}\varepsilon_{\rm B,-2}^{\frac{p+2}{2(p+4)}}n_2^{\frac{p+6}{2(p+4)}}\beta_{-1}^{\frac{p+6}{p+4}}R_{17}^{\frac{2}{p+4}}\ ,
    \label{eq:nu_a}
\end{align}
and the peak luminosity is given by
\begin{align}
\left(\nu L_\nu\right)_{\rm 1st}&\overset{p=2.5}{\simeq}8.9\times10^{35}{\,\rm erg\,s^{-1}\,}\bar{\varepsilon}_{\rm e,-1}^{\frac{7}{p+4}}\varepsilon_{\rm B,-2}^{\frac{3p+5}{2(p+4)}}
	\nonumber\\
&\,\,\,\,\,\,n_2^{\frac{3p+19}{2(p+4)}}R_{17}^{\frac{2p+15}{p+4}}\beta_{-1}^{\frac{3p+19}{p+4}}\left(\frac{\Omega}{4\pi}\right)\ .
\end{align}

Once both peak frequency and flux are obtained, we can estimate the radius of the outflow and the CNM density at the location:
\begin{align}
R&\overset{p=2.5}{\simeq}1.4\times10^{16}{\,\rm cm\,}\bar{\varepsilon}_{\rm e,-1}^{-\frac{1}{2p+13}}\varepsilon_{\rm B,-2}^{\frac{1}{2p+13}}
	\nonumber\\
&\,\,\,\,\,\,\nu_{\rm a,6GHz}^{-\frac{3p+19}{2p+13}}(\nu L_\nu)_{\rm 1st,38}^{\frac{p+6}{2p+13}}\left(\frac{\Omega}{4\pi}\right)^{-\frac{p+6}{2p+13}}\ ,
	\\
n&\overset{p=2.5}{\simeq}1.9\times10^4{\,\rm cm^{-3}\,}\bar{\varepsilon}_{\rm e,-1}^{-\frac{8}{2p+13}}\varepsilon_{\rm B,-2}^{-\frac{2p+5}{2p+13}}
	\nonumber\\
&\,\,\,\,\,\,\beta_{-1}^{-2}\nu_{\rm a,6GHz}^{\frac{2(2p+15)}{2p+13}}(\nu L_\nu)_{\rm 1st,38}^{-\frac{4}{2p+13}}\left(\frac{\Omega}{4\pi}\right)^{\frac{4}{2p+13}}\ .
\end{align}
For a freely-expanding outflow, using $R=\beta ct$, one obtains 
\begin{align}
\beta&\overset{p=2.5}{\simeq}0.056\,\bar{\varepsilon}_{\rm e,-1}^{-\frac{1}{2p+13}}\varepsilon_{\rm B,-2}^{\frac{1}{2p+13}}
	\nonumber\\
&\,\,\,\,\,\,\nu_{\rm a,6GHz}^{-\frac{3p+19}{2p+13}}(\nu L_\nu)_{\rm 1st,38}^{\frac{p+6}{2p+13}}\left(\frac{\Omega}{4\pi}\right)^{-\frac{p+6}{2p+13}}t_{\rm 100day}^{-1}\ ,
	\\
n&\overset{p=2.5}{\simeq}6.1\times10^4{\,\rm cm^{-3}\,}\bar{\varepsilon}_{\rm e,-1}^{-\frac{6}{2p+13}}\varepsilon_{\rm B,-2}^{-\frac{2p+7}{2p+13}}
	\nonumber\\
&\,\,\,\,\,\,\nu_{\rm a,6GHz}^{\frac{2(5p+34)}{2p+13}}(\nu L_\nu)_{\rm 1st,38}^{-\frac{2(p+8)}{2p+13}}\left(\frac{\Omega}{4\pi}\right)^{\frac{2(p+8)}{2p+13}}t_{\rm 100day}^{2}\ ,
\end{align}
where we normalize time by $100\,\rm days$.
Note this is practically the same calculation as the equipartition analysis with a correction for the deep-Newtonian phase \citep{Matsumoto&Piran2021b}.

\begin{figure}
\begin{center}
\includegraphics[width=85mm, angle=0,bb=0 0 278 265]{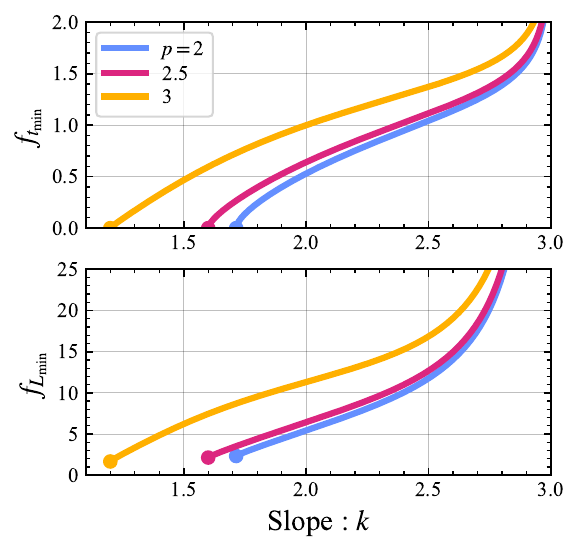}
\caption{The correction factors for the timescale and luminosity of the minimum in the light curve. They are defined by Eqs.~\eqref{eq:f_t_min} and \eqref{eq:f_L_min} and used in Eqs.~\eqref{eq:t_min} and \eqref{eq:L_min}. By definition, these factors are defined for $k>\frac{12}{p+5}$ (dots at the base of each curve).}
\label{fig:f}
\end{center}
\end{figure}

After producing the first radio peak at $t_{\rm 1st}$, the freely coasting outflow reaches the Bondi radius and if the CNM profile is steep enough, the light curve has a minimum  at 
\begin{align}
t_{\rm min}\simeq f_{t_{\rm min}}\Rb/v\simeq390{\,\rm day\,}\beta_{-1}^{-1}R_{\rm B,17}f_{
t_{\rm min}}\ ,
	\label{eq:t_min}
\end{align}
where we introduced a correction factor $f_{t_{\rm min}}$ representing a small shift of the minimum time from the Bondi-radius crossing time, $\Rb/v$, due to the smooth transition from the CNM to ISM of the density profile. This factor is obtained straightforwardly by the derivative of the optically-thin luminosity\footnote{Instead of Eq.~\eqref{eq:L_thin}, we use an exact form of the luminosity, which is given by $\left(\nu L_{\nu}\right)_{\rm thin}\propto n^{({p+1})/{4}}N(R)$, where $n$ and $N(R)$ are given by Eqs.~\eqref{eq:profile} and \eqref{eq:Ntot}, respectively.}:
\begin{eqnarray}
f_{t_{\rm min}}&=\left[\frac{k(p+1)-24}{24}+\sqrt{\left(\frac{k(p+1)-24 }{24}\right)^2 +\frac{k(p+5)-12}{4(3-k)}}\right]^{\frac{1}{k}}\ .
\label{eq:f_t_min}
\end{eqnarray}
The top panel of Fig.~\ref{fig:f} depicts the behavior of the correction factor for different $k$ and $p$. While its impact is within an order of unity, it could be important to estimate the size of the Bondi radius.
The density profile flattens outside of the Bondi radius. Once the outflow passes this radius, the luminosity increases, rising asymptotically as $ L_\nu \propto t^3$ (see Eq.~\ref{eq:L_thin} with $k=0$). Therefore, the luminosity has a minimal value at $t_{\rm min}$:
\begin{align}
\left(\nu L_{\nu}\right)_{\rm min}&\overset{p=2.5}{\simeq}1.8\times10^{37}{\,\rm erg\,s^{-1}\,}\bar{\varepsilon}_{\rm e,-1}\varepsilon_{\rm B,-2}^{\frac{p+1}{4}}
	\nonumber\\
&n_{\rm ISM,2}^{\frac{p+5}{4}}R_{\rm B,17}^3\beta_{-1}^{\frac{p+5}{2}}\nu_{\rm 6GHz}^{\frac{3-p}{2}}\left(\frac{\Omega}{4\pi}\right)f_{L_{\rm min},1}\ ,
	\label{eq:L_min}
\end{align}
here we replaced the density with that of the ISM which is constant outside of $\Rb$, and introduced an additional correction factor $f_{L_{\rm min}}$ to take into account for the underestimation by Eq.~\eqref{eq:L_thin} discussed earlier as well as a small shift of $t_{\rm min}$ from $\Rb/v$. This factor is given by
\begin{align}
f_{L_{\rm min}}=f_{t_{\rm min}}^3(1+f_{t_{\rm min}}^{-k})^{\frac{p+1}{4}}\left(1+\frac{3}{3-k}f_{t_{\rm min}}^{-k}\right)\ ,
    \label{eq:f_L_min}
\end{align}
and its dependence on $p$ and $k$ is depicted in the bottom panel of Fig.~\ref{fig:f}. 
Note that obtaining these correction factors, which depend on $p$ and $k$, is part of the fitting procedure of the data to the model. 

The radio light curve rises until the outflow starts decelerating. This occurs when the swept-up mass becomes comparable to the outflow's original mass at  
\begin{align}
R_{\rm dec}&\simeq\left(\frac{3M_{\rm ej}}{4\pi \MP n_{\rm ISM}}\right)^{1/3}\simeq6.6\times10^{17}{\,\rm cm\,}M_{\rm ej,-1}^{1/3}n_{\rm ISM,2}^{-1/3}\ ,
\end{align}
where $M_{\rm ej,-1} = M_{\rm ej}/(0.1 \Msun)$.
Therefore, the radio light curve has a second peak at 
\begin{align}
t_{\rm 2nd}&\simeq R_{\rm dec}/v\simeq2500{\,\rm day\,}M_{\rm ej,-1}^{1/3}n_{\rm ISM,2}^{-1/3}\beta_{-1}^{-1}\ ,
	\label{eq:t_dec}
\end{align}
whose luminosity is given by
\begin{align}
\left(\nu L_\nu\right)_{\rm 2nd}\overset{p=2.5}{\simeq}&5.2\times10^{38}{\,\rm erg\,s^{-1}\,}\bar{\varepsilon}_{\rm e,-1}\varepsilon_{\rm B,-2}^{\frac{p+1}{4}}
	\nonumber\\
&n_{\rm ISM,2}^{\frac{p+1}{4}}M_{\rm ej,-1}\beta_{-1}^{\frac{p+5}{2}}\nu_{\rm 6GHz}^{\frac{3-p}{2}}\left(\frac{\Omega}{4\pi}\right)\ .
	\label{eq:L_2nd}
\end{align}
After the second peak the outflow decelerates as $\beta\propto t^{-3/5}$ and $R\propto t^{2/5}$, and the radio luminosity declines as $\nu L_{\nu}\propto t^{-{3(p+1)}/{10}}$. Note that this peak may be the only peak if the inner density profile is sufficiently shallow.

The evolution of the luminosity $\nu L_\nu\propto t^3$, between the minimum and the second peak gives an interesting closure relation between the time and luminosity of the minimal and second peak:
\begin{align}
\frac{\left(\nu L_\nu\right)_{\rm 2nd}}{\left(\nu L_\nu\right)_{\rm min}}=\left(\frac{t_{\rm 2nd}}{t_{\rm min}}\right)^3\ ,
\end{align}
However, this may not strictly hold because an actual light curve evolves smoothly between the minimum and the second peak and flattens around both the minimum and peak, as shown in our numerical results (see the right panel of Fig.~\ref{fig:lc_para}). In particular, when the Bondi and deceleration radii are close, the light curve rises slowly.

\subsection{The diversity of radio light curves}\label{sec:diversity}
Given the radio light curve's basic nature, we now explore its parameter dependence. Here we focus on three key parameters: the slope of the CNM profile $k$, the ISM density $n_{\rm ISM}$, and the ejecta mass $M_{\rm ej}$.

\begin{figure*}
\begin{center}
\includegraphics[width=180mm, angle=0,bb=0 0 776 211]{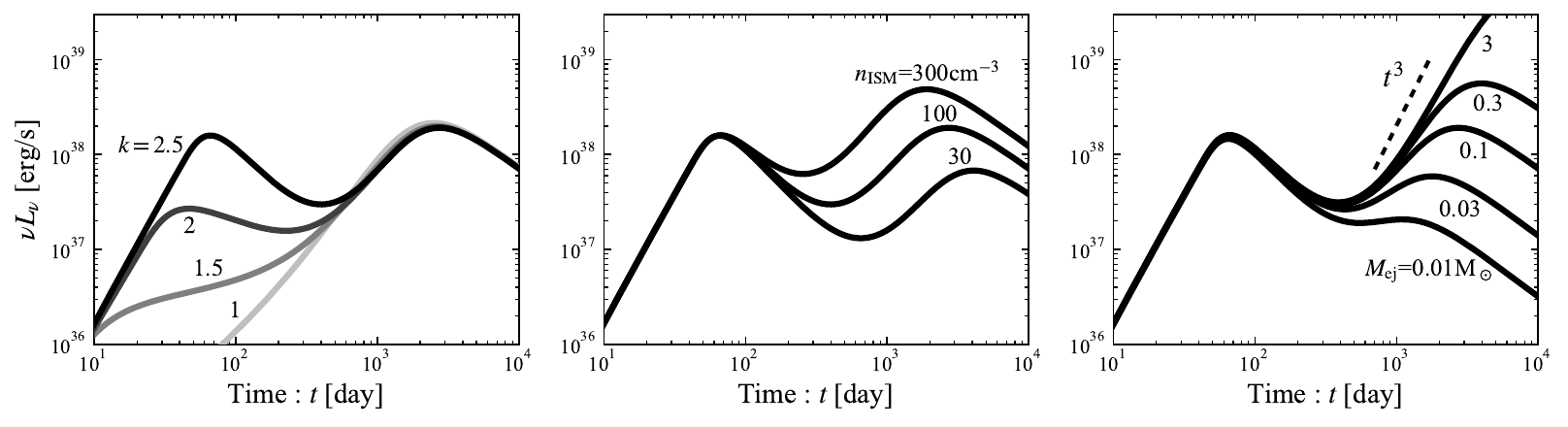}
\caption{Light curves for different slopes of the CNM density profile $k$, ISM densities $n_{\rm ISM}$, and ejecta masses $M_{\rm ej}$, from left to right. The density profiles adopted in the left panel are shown in Fig.~\ref{fig:profile}. In the middle panel, we adopt a different functional form of the density profile: Eq.~\eqref{eq:profile2} rather than Eq.~\eqref{eq:profile}.}
\label{fig:lc_para}
\end{center}
\end{figure*}

Fig.~\ref{fig:lc_para} depicts light curves for different $k$, $n_{\rm ISM}$, and $M_{\rm ej}$. The CNM density profile affects only the first peak of the radio light curve. Shallower profiles make the first peak earlier and less luminous because the CNM density is lower for a smaller $k$ value with a fixed Bondi radius and a fixed ISM density. In particular, for very shallow slopes of $k\lesssim{12}/{(p+5)}$, the first peak disappears and the light curve rises monotonically (see  Eq.~\eqref{eq:L_thin2}). Interestingly, for slopes expected for the Bondi accretion, $k\simeq 1.5$, we cannot detect a bright early-time radio flare at $\sim100\,\rm days$. This may explain why TDEs with early radio detection always have steep density profiles \cite[e.g.,][]{Alexander+2020}. A slow rise of the light curve of AT2019azh \citep{Goodwin+2022,Sfaradi+2022} may also be explained by the shallow density profile. We note that the minimum appears earlier for smaller $k$ corresponding to the behavior of the correction factor $f_{t_{\rm min}}$ in the top panel of Fig.~\ref{fig:f}.

The middle and right panels of Fig.~\ref{fig:lc_para} depict light curves obtained by varying $n_{\rm ISM}$ and $M_{\rm ej}$, respectively. As long as the deceleration radius is larger than the Bondi radius, both parameters impact the light curves only after the minimum. In the middle panel, we adopt a different parameterization for the density profile than Eq.~\eqref{eq:profile}
\begin{align}
n(R)=100{\,\rm cm^{-3}\,}\left(\frac{R}{10^{17}{\,\rm cm}}\right)^{-k}+n_{\rm ISM}\ ,
    \label{eq:profile2}
\end{align}
so that we have the same light curve around the first peak. With this parameterization, the Bondi radius or equivalently the radius at which the CNM and ISM densities are comparable is given by $\Rb=10^{17}{\,\rm cm}\left(\frac{100{\,\rm cm^{-3}}}{n_{\rm ISM}}\right)^{1/k}$.

Larger ISM densities give brighter radio signals due to the larger number of emitting particles, while shortening the timescale, resulting in earlier minimum and second peak. The ejecta mass changes the deceleration timescale. A larger ejecta mass also gives a brighter second peak because more mass is emitting during the peak. This behavior is similar to the one of radio afterglows of binary neutron star mergers \citep[e.g.,][]{Nakar&Piran2011}.

\subsection{A comparison with individual events}
\label{sec:fit}
To demonstrate the model, we compare it to a few observed late-time radio flares. We emphasize that due to a lack of sufficient data, we do not attempt to carry out a detailed ``best fit" procedure, and our aim is just to demonstrate the potential of this model. Before taking a closer look at individual events, we empathize two points: First, after the first peak, the outflow is optically thin to SSA at the observed frequency, and any variation in the light curve above this frequency should be achromatic. Second, once the emission becomes optically thin, the light curve cannot rise steeper than $\propto t^3$ unless the external density profile increases. While the current data is still limited, we notice that these points are satisfied by many events observed so far. Note, however, that some,  like AT 2018hyz, in which the light curve rises like $\propto t^5$, cannot be explained by this model and are better explained by a relativistic off-axis jet.

\begin{figure}
\begin{center}
\includegraphics[width=85mm, angle=0,bb=0 0 285 214]{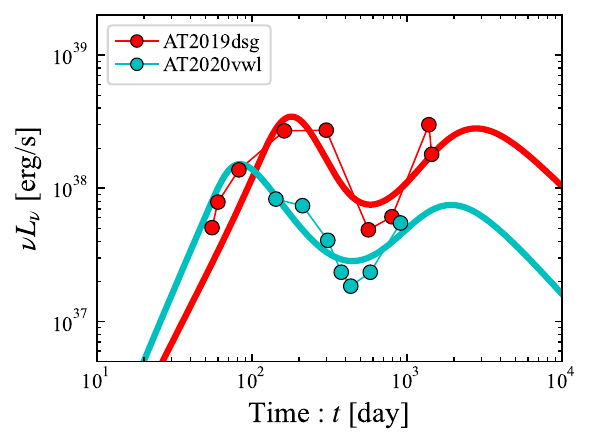}
\caption{Light curves of AT2019dsg and AT2020vwl at $6\,\rm GHz$, which are among the best late-time covered TDEs and possible light curves based on our model (see Table ~\ref{table:parameter_fit} for parameters). AT2019dsg shows a transition, at the first peak $\sim200\,\rm days$, from optically thick to optically thin spectrum at the observed frequency \citep{Stein+2021,Cendes+2021b}. The spectrum of AT2020vwl may hint at a similar transition at the first two epochs \citep{Goodwin+2023b}.}
\label{fig:lc_obs}
\end{center}
\end{figure}

Fig.~\ref{fig:lc_obs} depicts the observed and model light curves at $6\,\rm GHz$ for two of the best observed late-time flares in TDEs: AT2019dsg \citep{Stein+2021,Cendes+2021b,Cendes+2023} that shows a double peak structure and AT2020vwl \citep{Goodwin+2023b,Goodwin+2023Atel} that show a clear minimum. The outflow is assumed to be launched at the time of discovery of the TDE.\footnote{See \cite{Matsumoto+2022} for a discussion of the origin of the outflow in AT2019dsg. We also showed that the radio emission of AT2019dsg cannot be explained by an off-axis jet in \cite{Matsumoto&Piran2023}.} The adopted parameters and corresponding density profiles (For AT2019dsg, we adopt a broken power-law function for the CNM profile to mimic the result by the equipartition analysis) are shown in Table~\ref{table:parameter_fit} and Fig.~\ref{fig:profile2}. Again, we stress that these parameters are not obtained by exploring the entire parameter space. We find that both events are reasonably reproduced by typical parameter values of $p\simeq2.5$, $\varepsilon_{\rm e}\simeq0.1$, $\varepsilon_{\rm B}\simeq0.01$, and $\beta\simeq0.1$ while we adopt the observationally obtained values for $p$. We also confirm that the radio spectra are also reasonably reproduced by our model.

\begin{figure}
\begin{center}\includegraphics[width=85mm, angle=0,bb=0 0 285 219]{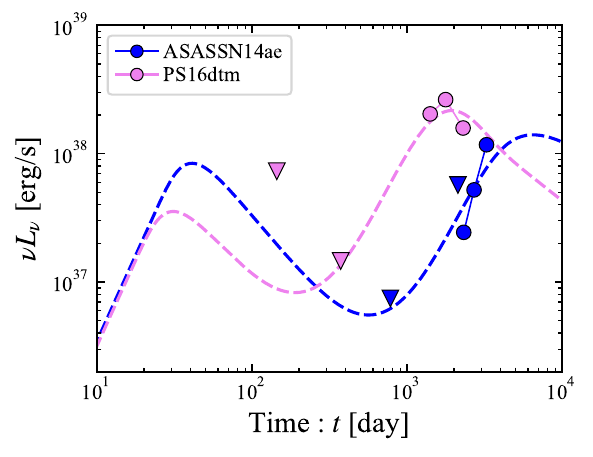}
\caption{Examples of $6\,\rm GHz$ light curves of TDEs with late-time scarce observations: PS16dtm and ASSASN14ae. Both events show strong early upper limits. PS16dtm shows a late-time maximum, and ASSASN14ae shows a late-time rapid rise. The dashed lines show tentative light curves that follow from our model (see Table ~\ref{table:parameter_fit} for parameters). Even though the fits are not unique, valuable information about the source can be obtained from the data.}
\label{fig:lc_obs2}
\end{center}
\end{figure}

As an example for TDEs that have a flare with limited late-time data, Fig.~\ref{fig:lc_obs2} depicts two events taken from \cite{Cendes+2023}: PS16dtm showing a late-time optically thin maximum, and ASSASN14ae showing a fast rise at $\simeq2000\,\rm days$. Clearly, these events do not have enough data to obtain a unique fit. Still, we present in Fig.~\ref{fig:lc_obs2} a tentative fit whose parameters are given in Table ~\ref{table:parameter_fit}. We demonstrate in the following that some information on the source can be obtained even with such minimal data.

\begin{figure}
\begin{center}
\includegraphics[width=85mm, angle=0,bb=0 0 276 241]{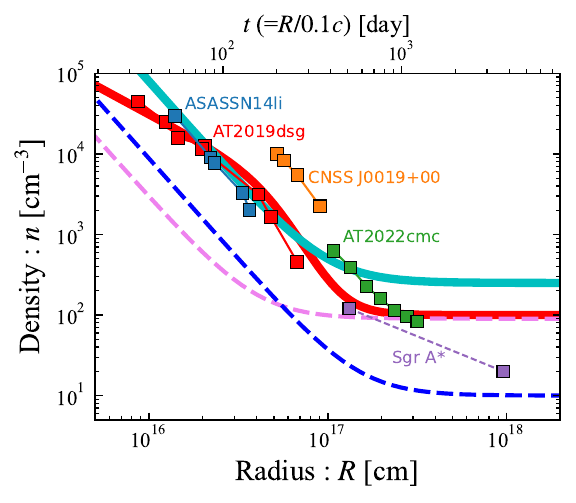}
\caption{
Density profiles adopted to depict the light curves of individual objects in Figs.~\ref{fig:lc_obs} and ~\ref{fig:lc_obs2}. The color scheme is the same as in those figures. Squares represent density profiles obtained by the equipartition analysis of the early light curve (the same as those shown in Fig.~\ref{fig:profile}).}
\label{fig:profile2}
\end{center}
\end{figure}

\begin{table*}
\begin{center}
\caption{Fitting parameters for individual events in Fig.~\ref{fig:lc_obs}.}
\label{table:parameter_fit}
\begin{tabular}{lcccccccc}
\hline
Event&$p$&$\varepsilon_{\rm e}$&$\varepsilon_{\rm B}$&$M_{\rm ej}$&$\beta_0$&$n_{\rm ISM}$&$\Rb$&$k$\\
&&&&[$\Msun$]&&[cm$^{-3}$]&[$10^{17}$cm]&\\
\hline
AT2019dsg&2.7&0.2&0.02&0.1&0.1&100&1&1.2 \& 4.5\\
AT2020vwl&3&0.15&0.01&0.1&0.1&250&1&2.5\\
\hline
PS16dtm&2.1&0.13&0.01&0.04&0.08&90&0.4&2.5\\
ASASSN14ae&2.2&0.08&0.01&0.2&0.1&10&1.5&2.5\\
\hline
\end{tabular}
\end{center}
\end{table*}

\subsection{Parameters Inference }\label{sec:parameter}
The emission after the first peak is described by an optically thin synchrotron. Thus, the radio reflects the density profile of the surrounding medium and the outflow dynamics. Clearly, a good fit to the whole light curve directly provides the density profile of the CNM and the ISM surrounding the SMBH. Such curves are shown in Fig.~\ref{fig:profile2} for the events discussed earlier. 

We can constrain the parameters characterizing the radio light curve and the density even when we do not have the complete light-curve data, but the minimum or the second peak is well identified. When a radio light curve has a minimum at $t_{\rm min}$ with $\left(\nu L_\nu\right)_{\rm min}$ and given the ejecta velocity, for example from earlier observations, these observables give estimates of the Bondi radius and the ISM density:
\begin{align}
\Rb&\simeq {vt_{\rm min}}/{f_{t_{\rm min}}}\simeq7.8\times10^{16}{\,\rm cm\,}\beta_{-1}t_{\rm min,300day}f_{t_{\rm min}}^{-1}\ ,
	\label{eq:R_b2}\\
n_{\rm ISM}&\overset{p=2.5}{\simeq} 1.1\times10^2{\,\rm cm^{-3}\,}\bar{\varepsilon}_{\rm e,-1}^{-\frac{4}{p+5}}\varepsilon_{\rm B,-2}^{-\frac{p+1}{p+5}}\beta_{-1}^{-\frac{2(p+11)}{p+5}}
	\nonumber\\
&\,\,\,\,\,\,\nu_{\rm 6GHz}^{\frac{2(p-3)}{p+5}}(\nu L_\nu)_{\rm min,37}^{\frac{4}{p+5}}\left(\frac{\Omega}{4\pi}\right)^{-\frac{4}{p+5}}t_{\rm min,300day}^{-\frac{12}{p+5}}f_{t_{\rm min}}^{\frac{12}{p+5}}f_{L_{\rm min},1}^{-\frac{4}{p+5}}\ ,
	\label{eq:n_ISM}
\end{align}
where we used Eqs.~\eqref{eq:t_min} and \eqref{eq:L_min}, and the minimum timescale is normalized by $300\,\rm days$.

\begin{figure}
\begin{center}
\includegraphics[width=85mm, angle=0,bb=0 0 277 242]{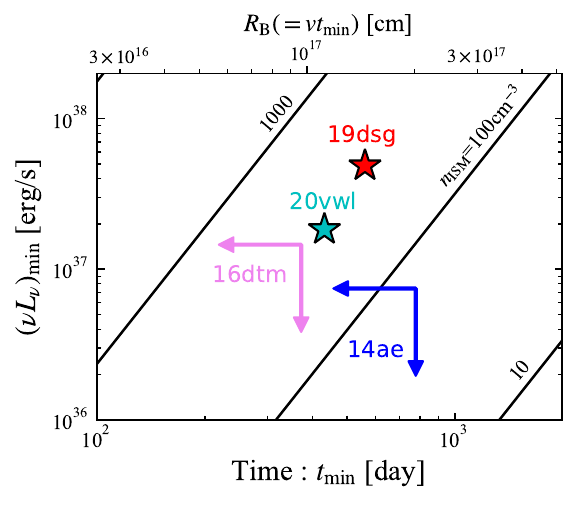}
\caption{Observationally identified (stars) or constrained (arrows) minima in radio light curves. Diagonal lines represent contours of fixed ISM density. The top horizontal axis shows the Bondi radius corresponding to the minimum timescale. These contours and axis are obtained for parameters of $p=2.5$, $\varepsilon_{\rm e}=0.1$, $\varepsilon_{\rm B}=0.01$, $\beta=0.1$, $f_{t_{\rm min}}=1$, and $f_{L_{\rm min}}=10$.}
\label{fig:tL_min}
\includegraphics[width=85mm, angle=0,bb=0 0 292 242]{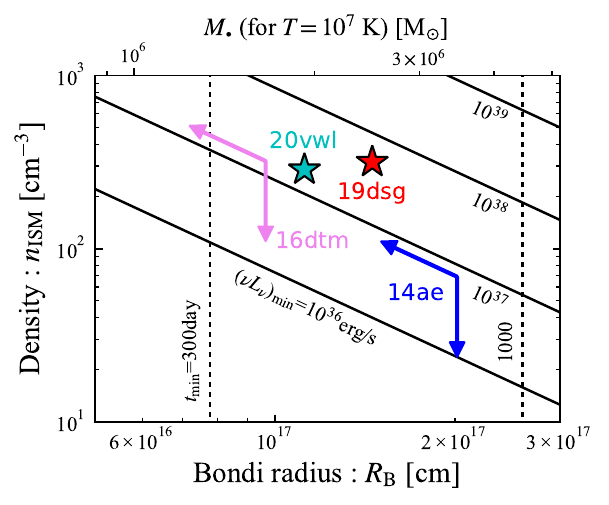}
\caption{The same as Fig.~\ref{fig:tL_min} with the Bondi radius and the ISM density as horizontal and vertical axes, respectively. The top horizontal axis shows the corresponding BH mass estimated for $T=10^7\,\rm K$.}
\label{fig:tL_min2}
\end{center}
\end{figure}

Figs.~\ref{fig:tL_min} and \ref{fig:tL_min2} demonstrate how these relations constrain  $\Rb$ and $n_{\rm ISM}$ assuming all events share the same parameters of $p=2.5$, $\varepsilon_{\rm e}=0.1$, $\varepsilon_{\rm B}=0.01$, and $\beta=0.1$. Realistically, each event has different parameter values and they cannot be put in the same figure. In Fig.~\ref{fig:tL_min}, ISM density contours are drawn along with observed events. Identifying the minimum is possible only for AT2019dsg and AT2020vwl, which show a secondary rise. Other events showing a rising light curve with an optically thin SED put only upper limits on the timescale and luminosity of the minima. In Fig.~\ref{fig:tL_min2}, we recast the relations and draw minimum timescale and luminosity contours. Intriguingly, constraining the Bondi radius allows us to infer the BH mass up to the uncertainty of the ISM sound velocity (or, equivalently, the ISM temperature). As an example, we add an axis for the BH mass at the top of the figure, assuming $T=10^7\,\rm K$. For the observed events, the BH mass is constrained to be $\lesssim10^6\,\Msun$, similar to expected values for typical TDEs \citep{Ryu+2020,Hammerstein+2023,Yao+2023}.  

\begin{figure}
\begin{center}
\includegraphics[width=85mm, angle=0,bb=0 0 285 242]{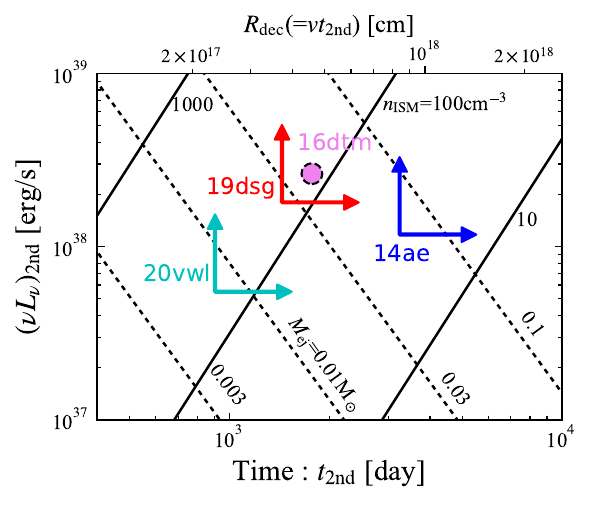}
\caption{The same as Fig.~\ref{fig:tL_min} but for the second peak. Note that the data for PS16dtm is rather scarce, and hence, the estimates are less certain.}
\label{fig:tL_max}
\includegraphics[width=85mm, angle=0,bb=0 0 276 243]{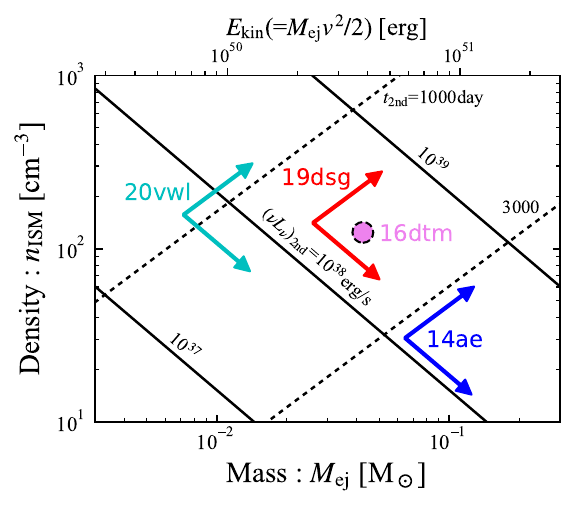}
\caption{The same as Fig.~\ref{fig:tL_max2} but for the second peak, which constrains the ejecta mass and ISM density. The top horizontal axis shows the corresponding kinetic energy for $\beta=0.1$. Note that the data for PS16dtm is rather scarce, and hence, the estimates are less certain.}
\label{fig:tL_max2}
\end{center}
\end{figure}

The timescale and luminosity of the second peak (or, more generally, a radio peak caused by a deceleration of an optically-thin outflow) constrain the ejecta mass and ISM density:
\begin{align}
M_{\rm ej}&\overset{p=2.5}{\simeq} 3.9\times10^{-2}{\,\Msun\,}\bar{\varepsilon}_{\rm e,-1}^{-\frac{4}{p+5}}\varepsilon_{\rm B,-2}^{-\frac{p+1}{p+5}}\beta_{-1}^{-\frac{p-7}{p+5}}
	\nonumber\\
&\,\,\,\,\,\,\nu_{\rm 6GHz}^{\frac{2(p-3)}{p+5}}(\nu L_\nu)_{\rm 2nd,39}^{\frac{4}{p+5}}\left(\frac{\Omega}{4\pi}\right)^{-\frac{4}{p+5}}t_{\rm 2nd,1000day}^{\frac{3(p+1)}{p+5}}\ ,
    \label{eq:Mej}\\
n_{\rm ISM}&\overset{p=2.5}{\simeq} 6.3\times10^2{\,\rm cm^{-3}\,}\bar{\varepsilon}_{\rm e,-1}^{-\frac{4}{p+5}}\varepsilon_{\rm B,-2}^{-\frac{p+1}{p+5}}\beta_{-1}^{-\frac{2(p+11)}{p+5}}
	\nonumber\\
&\,\,\,\,\,\,\nu_{\rm 6GHz}^{\frac{2(p-3)}{p+5}}(\nu L_\nu)_{\rm 2nd,39}^{\frac{4}{p+5}}\left(\frac{\Omega}{4\pi}\right)^{-\frac{4}{p+5}}t_{\rm 2nd,1000day}^{-\frac{12}{p+5}}\ .
\end{align}
The second equation is basically the same as Eq.~\eqref{eq:n_ISM} but for the timescale and luminosity of the second peak. These relations are shown in Figs.~\ref{fig:tL_max} and \ref{fig:tL_max2} similarly to the minima discussed earlier. A relatively secure identification of the second peak is possible only for  PS16dtm. Other events showing a rise give only lower limits on $t_{\rm 2nd}$ and $\left( \nu L_\nu\right)_{\rm 2nd}$. 

The second peak allows us to constrain the mass and kinetic energy of the radio-emitting outflow. The latter is estimated by using Eq.~\eqref{eq:Mej}:
\begin{align}
E_{\rm kin}&\overset{p=2.5}{\simeq}\frac{1}{2}M_{\rm ej}v^2\simeq3.5\times10^{50}{\,\rm erg\,}\bar{\varepsilon}_{\rm e,-1}^{-\frac{4}{p+5}}\varepsilon_{\rm B,-2}^{-\frac{p+1}{p+5}}\beta_{-1}^{\frac{p+17}{p+5}}
	\nonumber\\
&\,\,\,\,\,\,\nu_{\rm 6GHz}^{\frac{2(p-3)}{p+5}}(\nu L_\nu)_{\rm 2nd,39}^{\frac{4}{p+5}}\left(\frac{\Omega}{4\pi}\right)^{-\frac{4}{p+5}}t_{\rm 2nd,1000day}^{\frac{3(p+1)}{p+5}}\ .
	\label{eq:Ekin}
\end{align}
Within our approximation of a single velocity outflow Eqs.~\eqref{eq:Mej} and \eqref{eq:Ekin} estimate the total mass and energy. However,  the radio luminosity depends strongly on the outflow velocity (see Eq. ~\ref{eq:L_2nd}). Thus, a slower outflow component will not contribute much to the radio signal. As such, generally, depending on the velocity structure of the outflow, Eqs.~\eqref{eq:Mej} and \eqref{eq:Ekin} may provide only lower limits on the mass and energy. 

For PS16dtm, the ejecta mass is constrained to $\lesssim 0.1\,\Msun$, which is significantly smaller than that required for the reprocessing outflow model for optical emissions (e.g., \citealt{Metzger&Stone2016}, but see \citealt{Matsumoto&Piran2021} for the mass budget of optical TDE models), and that of the unbound debris \citep[e.g.,][]{Krolik+2016}. However, one has to recall that this is just a lower limit and a slower component may exist.

It is important to stress that these last estimates of the ejected mass and energy are quite general. For example, it has been suggested that the late radio flares arise from delayed outflows \citep{Cendes+2023,Teboul&Metzger2023,Lu+2023}. Since, in most cases, the emission is optically thin (at least the high-frequency part of the radio spectra), 
regardless of the launching time, these mass and energy estimates should be valid. Even these current mass estimates and lower limits already constrain some scenarios for producing those delayed outflows.

\section{Summary}\label{sec:summary}
Late-time TDE radio flares \citep{Horesh+2021b, Cendes+2022b,Goodwin+2022,Perlman+2022,Sfaradi+2022,Cendes+2023,Goodwin+2023b,Somalwar+2023c,Sfaradi+2024,ZhangFabao+2024,Christy+2024} are an intriguing part of the TDE puzzle.  Some cases, notably AT2018hyz, show a very steep rise in late-time. Those late radio flares arise from off-axis relativistic jets that are slowing down and whose beamed emission is coming into our line of sight \citep{Matsumoto&Piran2023,Beniamini+2023c,Sfaradi+2024}. However, off-axis relativistic jets cannot explain all events and in particular \cite{Matsumoto&Piran2023} have shown that the observed radio light curve of AT2019dsg is incompatible with this interpretation.

Here, we propose a novel model to explain some of the late-time radio flares. In this model, the density profile around the SMBH flattens outside of the Bondi radius. An outflow expanding at a constant velocity can naturally produce the observed double-peak radio light curve (Fig.~\ref{fig:lc}) as seen for AT2019dsg and AT2020vwl. The first peak is produced by a transition from SSA-thick to SSA-thin optical depth (at the observed frequency) within a decreasing density profile as suggested by previous observations \citep[e.g.,][]{Krolik+2016,Alexander+2016,Anderson+2020}. However, we found that 
depending on the slope of the density profile, the first peak may not always appear (see the left panel of Fig.~\ref{fig:lc_para}). In particular, the light curve rises monotonically for density profiles shallower than $R^{-3/2}$, which are expected for Bondi accretion. In cases when the first peak appears, the radio light curve declines until the outflow reaches the Bondi radius at $\sim10^{17}\,\rm cm$ (see Eq.~\ref{eq:R_bondi}). Beyond the Bondi radius, a constant density profile is expected. The light curve increases\footnote{Note that light curves rising faster than $t^3$, as seen for example in AT2018hyz, cannot be explained if the external density is constant.} asymptotically like $\propto t^3$, until the swept-up ISM mass becomes comparable to the outflow's mass, and the outflow starts decelerating. This leads to the second\footnote{This would be the only peak for a shallow external density profile.} peak.

We did not attempt to perform a detailed parameter fitting to the observed data, mainly because the data was relatively scarce for most events. Additionally, some of the approximations used in this work, particularly the single velocity approximation, may not be valid around the deceleration time. Still, we demonstrated that the model reasonably fits some of the observed late-time radio light curves (see Fig.~\ref{fig:lc_obs}). Within our model, the detection of the light curve minimum and the second maximum allows us to constrain the BH mass and ISM density and set lower limits on the ejecta mass and energy of the outflow if the outflow's velocity is known, for example, from an analysis of the first peak. For our sample, the estimated BH masses, $\lesssim3\times10^6\,\Msun$, are within a range of a typical SMBH mass (Fig.~\ref{fig:tL_min2}). The estimated ejecta mass (Fig.~\ref{fig:tL_max2}), which is $\gtrsim 0.01\,\Msun$, is consistent with the one expected from unbound debris \citep{Krolik+2016} or from the reprocessed outflow for optical emissions \citep{Metzger&Stone2016}. 

To summarize, we have outlined a model for the production of late-radio flares from TDEs and interpretation of their observations. Current data is, in practically all cases, insufficient, and as such, we compared it only to a simplified model that ignores the velocity distribution within the outflow. Further detailed observations, combined with generalization of the presented model including the outflow velocity distribution (which is possible at least in the unbound debris model) can rule out or confirm this model and in case it is acceptable provide new probes on the TDE enigma.

\begin{acknowledgements}
We thank Yvette Cendes for providing us with data of late-time radio flares. T.M. also thanks Keiichi Maeda and Kunihito Ioka for their useful comments. T.P. thanks the Yukawa Institute for Theoretical Physics at Kyoto University for hospitality while this work was initiated. This research is supported by the Hakubi project at Kyoto University, JSPS KAKENHI Grant Number 24K17088 (T.M.), Advanced ERC grant MultiJets, ISF grant 2126/22, and the Simons Collaboration on Extreme Electrodynamics of Compact Sources (T.P.).
\end{acknowledgements}

%\appendix

%#\bibliographystyle{mn2e}
\bibliographystyle{aasjournal}
%\bibliography{refs}
\bibliography{reference_matsumoto}

\end{document}